\documentclass[aps,amssymb,showpacs,twocolumn]{revtex4}
\usepackage{amssymb}
\usepackage{graphicx}
\usepackage{color}
\usepackage{textcomp}
\usepackage{ulem}
\begin{document}

\title{ Magnetic field effect on pion superfluid}
\author{Shijun Mao}
 \email{maoshijun@mail.xjtu.edu.cn}
\affiliation{School of Physics, Xi'an Jiaotong University, Xi'an, Shaanxi 710049, China}

\begin{abstract}
Magnetic field effect on pion superfluid phase transition is investigated in frame of a Pauli-Villars regularized NJL model. Instead of directly dealing with charged pion condensate, we apply the Goldstone's theorem (massless Goldstone boson $\pi^+$) to determine the onset of pion superfluid phase, and obtain the phase diagram in magnetic field, temperature, isospin and baryon chemical potential space. At weak magnetic field, it is analytically proved that the critical isospin chemical potential of pion superfluid phase transition is equal to the mass of $\pi^+$ meson in magnetic field. The pion superfluid phase is retarded to higher isospin chemical potential, and can survive at higher temperature and higher baryon chemical potential under external magnetic field.
\end{abstract}

\date{\today}
\maketitle

The study of QCD at finite isospin density and the corresponding pion superfluid phase attracts much attention due to its relation to the investigation of compact stars, isospin asymmetric nuclear matter, and heavy-ion collisions at intermediate energies. On numerical side, while there is not yet precise lattice results at finite baryon density due to the Fermion sign problem, it is in principle no problem to do lattice simulation at finite isospin density~\cite{lqcd1,lqcd2,lqcd3}. On analytical side, effective models such as Nambu--Jona-Lasinio model (NJL), linear sigma model and chiral perturbation theory have been widely used to investigate pion superfluid phase structure~\cite{model1,model2,model3,model4,model5,model6,model7,model8,model9,model10,model11,model12,model13,
model14,model15,model16,model17,model18,model19,model20,model21,model22,model23}. There are two equivalent criteria for the critical point of pion superfluid phase transition, the non-vanishing charged pion condensate and the massless $\pi^+$ meson, which correspond to the spontaneous breaking of isospin symmetry and the Goldstone boson, respectively, guaranteed by the Goldstone's theorem~\cite{gold1,gold2}. With vanishing temperature, the critical isospin chemical potential $\mu_I^c$ is the pion mass in vacuum $m_\pi$. When $\mu_I>m_\pi$, the $u$ quark and $\bar d$ quark form coherent pairs and condensate, and the system enters the pion superfluid phase~\cite{lqcd1,lqcd2,lqcd3,model1,model2,model3,model4,model5,model6,model7,model8,model9,model10,model11,model12,model13,
model14,model15,model16,model17,model18,model19,model20,model21,model22,model23}. At hadron level, in the normal phase $(\mu_I < m_\pi)$ without charged pion condensate, different pion modes explicitly show the mass splitting according to their isospin, with $m_{\pi^\pm}=m_\pi\mp \mu_I$ and $m_{\pi^0}=m_\pi$. As $\mu_I=\mu_I^c=m_\pi$, the excitation of $\pi^+$ meson is free with zero momentum, which indicates the onset of pion superfluid phase~\cite{lqcd2,model9,model12,model13,model14,model21}. Inside the pion superfluid phase $(\mu_I \geq m_\pi)$, $\pi^+$ meson keeps massless as the Goldstone mode~\cite{lqcd2,model9,model12,model13,model14,model21}.

Recently, the magnetic properties of QCD matter become important. For instance, a certain class of neutron stars (magnetars) exhibits intense magnetic fields of strengths up to $10^{14-15}$ Gauss at the star surface and the field is expected to become stronger towards the star center, about $10^{18}$ Gauss~\cite{neutron1,neutron2}. However, the magnetic field effect on the pion superfluid is still an open question. The difficulty lies in the fact that the pion superfluid is a phase with charged pion condensate. It breaks both the isospin symmetry in the flavor space and the translational invariance in the coordinate space, and thus the Fourier transformation between coordinate and momentum spaces is not as simple as for neutral condensate or without magnetic field. LQCD simulations exhibit a sign problem at finite isospin chemical potential and magnetic field. By using a Taylor expansion in the magnetic field, it is reported that at vanishing temperature, the onset of pion condensate shifts to larger isospin chemical potential under magnetic fields~\cite{lqcdb1}, which is qualitatively consistent with the enhancement of the charged pion mass with growing magnetic fields~\cite{lqcdb2}. In the study of effective models, people also focus on the charged pion condensate but the interaction between the charged pion condensate and the magnetic field is simply neglected in Ref.~\cite{pib1,pib2} or taken into account by the Ginzburg-Landau approach assuming a tiny condensate in Ref.~\cite{pib3}.

In this paper, we will study the pion superfluid phase transition at finite magnetic field, temperature, isospin and baryon chemical potential in frame of a Pauli-Villars regularized NJL model, which is inspired by the Bardeen-Cooper-Shrieffer (BCS) theory and describes remarkablely well the quark pairing mechanisms and hadron mass spectra~\cite{njl1,njl2,njl3,njl4,njl5,zhuang}. Instead of directly dealing with charged pion condensate, we investigate the magnetic field effect on pion superfluid through its Goldstone mode $\pi^+$, determining the critical point of pion superfluid phase transition by the massless $\pi^+$ meson. Seriously taking into account the breaking of translational invariance for
charged particles, the pion propagators in terms of quark bubbles are analytically derived, and pion masses are solved. At weak magnetic field and vanishing temperature and baryon chemical potential, we analytically prove that the critical isospin chemical potential of pion superfluid phase transition is equal to the $\pi^+$ mass in magnetic field, the same as the vanishing magnetic field case~\cite{lqcd2,model9,model12,model13,model14,model21}. Under external magnetic field, the pion superfluid phase is shifted to higher isospin chemical potential, and can survive at higher temperature and higher baryon chemical potential.

The two-flavor NJL model is defined through the Lagrangian density in terms of quark fields $\psi$~\cite{njl1,njl2,njl3,njl4,njl5,zhuang}
\begin{equation}
\label{njl}
{\cal L} = \bar{\psi}\left(i\gamma_\nu D^\nu-m_0+\gamma_0 \mu\right)\psi+G \left[\left(\bar\psi\psi\right)^2+\left(\bar\psi i\gamma_5{\vec \tau}\psi\right)^2\right].
\end{equation}
Here the covariant derivative $D_\nu=\partial_\nu+iQ A_\nu$ couples quarks with electric charge $Q=diag (Q_u,Q_d)=diag (2e/3,-e/3)$ to the external magnetic field ${\bf B}=(0, 0, B)$ in $z$-direction through the potential $A_\nu=(0,0,Bx_1,0)$. The quark chemical potential $\mu
=diag\left(\mu_u,\mu_d\right)=diag\left(\mu_B/3+\mu_I/2,\mu_B/3-\mu_I/2\right)$ is a matrix in the flavor space, with $\mu_u$ and $\mu_d$
being the $u$- and $d$-quark chemical potentials and $\mu_B$ and
$\mu_I$ being the baryon and isospin chemical potentials. $G$ is the coupling constant in scalar and pseudo-scalar channels. At finite isospin chemical potential and magnetic field, the isospin symmetry $SU(2)_I$ is broken down to $U(1)_I$ symmetry, and the chiral symmetry $SU(2)_A$ is broken down to $U(1)_A$ symmetry. With the spontaneous breaking of chiral $U(1)_A$ symmetry and isospin $U(1)_I$ symmetry, the Goldstone mode reads $\pi^0$ meson and $\pi^+$ meson, respectively. $m_0$ is the current quark mass characterizing the explicit chiral symmetry breaking.

Corresponding to the symmetries and their spontaneous breaking, we have two order parameters, neutral chiral condensate $\langle\bar\psi\psi\rangle$ for chiral restoration phase transition and charged pion condensate $\langle\bar\psi\gamma_5\tau^1\psi\rangle$ for pion superfluid phase transition. Under magnetic fields, the charged pion condensate breaks both the isospin symmetry in the flavor space and the translational invariance in the coordinate space, and thus the Fourier transformation between coordinate and momentum spaces is not as simple as for neutral condensate or without magnetic field. In our current work, to avoid the complication and difficulty of dealing with charged pion condensate under magnetic field, we will start from the normal phase only with neutral chiral condensate and determine the critical point of pion superfluid phase transition by the appearance of Goldstone boson, massless $\pi^+$ meson. Physically, it is equivalent to define the phase transition by the order parameter (charged pion condensate) and Goldstone mode (massless $\pi^+$ meson), as guaranteed by the Goldstone's theorem~\cite{gold1,gold2,lqcd2,model9}.

In mean field approximation, the chiral condensate $\langle\bar\psi\psi\rangle$ or the dynamical quark mass $m_q=m_0-2G\langle\bar\psi\psi\rangle$ is controlled by the gap equation~\cite{rev1,rev2,rev3,rev4,rev5,rev6,rev7},
\begin{eqnarray}
\label{gap}
m_0&=&m_q(1-2GJ_1), \\
J_1 &=& 3\sum_{f,n}\alpha_n \frac{|Q_f B|}{2\pi} \int \frac{d p_3}{2\pi} \frac{1}{ E_f}  \\ &&\times \left[1-f(E_f+\mu_f)-f(E_f-\mu_f) \right], \nonumber
\end{eqnarray}
with the summation over all flavors and Landau energy levels, spin factor $\alpha_n=2-\delta_{n0}$, quark energy $E_f=\sqrt{p^2_3+2 n |Q_f B|+m_q^2}$, and Fermi-Dirac distribution function $f(x)=1/(e^{x/T}+1)$.

As quantum fluctuations above the mean field, mesons are constructed through quark bubble summations in the frame of random phase approximation~\cite{njl2,njl3,njl4,njl5,zhuang}. Taking into account of the interaction between charged mesons and magnetic fields, and generalizing our derivations in Ref.~\cite{rev7} to finite quark chemical potential, the meson propagator $D_M$ can be expressed in terms of the meson polarization function $\Pi_M$ with conserved Ritus momentum ${\bar k}$,
\begin{eqnarray}
D_M({\bar k})=\frac{G}{1-G\Pi_M({\bar k})}.
\end{eqnarray}
The meson pole mass $m_M$ is defined through the pole of the propagator at zero momentum,
\begin{eqnarray}
1-G\Pi_M(k_0=m_M)=0.
\label{pip}
\end{eqnarray}

\begin{figure}[hbt]
\centering
\includegraphics[width=7.5cm]{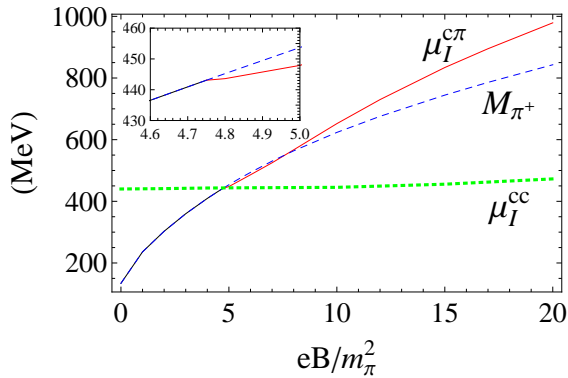}
\caption{Critical isospin chemical potential $\mu_I^{c\pi}$ (black and red solid lines) for pion superfluid phase transition, and $\mu_I^{cc}$ (green dotted line) for chiral restoration phase transition as a function of magnetic field at $T=\mu_B=0$. $\pi^+$ mass in magnetic field $M_{\pi^+}=m_{\pi^+}(B,T=\mu_B=\mu_I=0)$ is plotted in blue dashed line for reference.} \label{muicb}
\end{figure}

Based on the Goldstone's theorem for the spontaneous breaking of isospin symmetry, massless Goldstone mode $\pi^+$ exists in the pion superfluid phase. Therefore, the critical isospin chemical potential $\mu_I^{c\pi}$ for pion superfluid can be identified by the condition
\begin{eqnarray}
m_{\pi^+}(B,T,\mu_B,\mu_I^{c\pi})=0.
\end{eqnarray}
For the $\pi^+$ meson, we have 
\begin{eqnarray}
\Pi_{\pi^+}(k_0) &=& J_1+J_2(k_0),\\
J_2(k_0) &=& \sum_{n,n'} \int \frac{d p_3}{2\pi}\frac{j_{n,n'}(k_0)}{4E_n E_{n'}}\\
&&\times \big[\frac{f(-E_{n'}-\mu_u)- f(E_n-\mu_d)}{k_0+\mu_I+E_{n'}+E_n}\nonumber\\&&+\frac{f(E_{n'}-\mu_u)- f(-E_n-\mu_d)}{k_0+\mu_I-E_{n'}-E_n}\big],\nonumber\\
j_{n,n'}(k_0) &=& \left[{(k_0+\mu_I)^2/2}-n'|Q_u B|-n|Q_d B|\right]j^+_{n,n'} \nonumber\\
&&-2 \sqrt{n'|Q_u B|n|Q_d B|}\ j^-_{n,n'},
\end{eqnarray}
with the $u$-quark energy $E_{n'}=\sqrt{p^2_3+2 n' |Q_u B|+m_q^2}$ and $d$-quark energy $E_n=\sqrt{p^2_3+2 n |Q_d B|+m_q^2}$. The coefficients $j^{\pm}_{n,n'}$ are detailed derived in our previous work~\cite{rev7}. Note that the lowest-Landau-level term with $n=n'=0$ do not contribute to the polarization function with $j^{\pm}_{0,0}=0$. Because the spins of $u$ and $\bar d$ quarks at the lowest Landau level are aligned parallel to the magnetic field, but $\pi^+$ meson has spin zero. This leads to the heavy $\pi^+$ mass in magnetic field~\cite{rev7} and thus delays the pion superfluid in magnetic field (see the discussions of Fig.\ref{muicb}).

Because of the four-fermion interaction, the NJL model is not a renormalizable theory and needs regularization. The magnetic field does not cause extra ultraviolet divergence but introduces discrete Landau levels and anisotropy in momentum space. To guarantee the law of causality in anisotropic systems, we take into account the gauge invariant Pauli-Villars regularization scheme~\cite{rev6,rev7}. The three parameters in the NJL model, namely the current quark mass $m_0=5$ MeV, the coupling constant $G=3.44$ GeV$^{-2}$ and the Pauli-Villars mass parameter $\Lambda=1127$ MeV are fixed by fitting the chiral condensate $\langle\bar\psi\psi\rangle=-(250\ \text{MeV})^3$, pion mass $m_\pi=134$ MeV and pion decay constant $f_\pi=93$ MeV in vacuum with $T=\mu_B=\mu_I=0$ and $B=0$.

In Fig.\ref{muicb}, we plot the critical isospin chemical potential $\mu_I^{c\pi}$ (black and red solid lines) for pion superfluid phase transition as a function of magnetic field at $T=\mu_B=0$, which is determined by the condition of massless Goldstone boson $m_{\pi^+}(B,T=\mu_B=0,\mu_I^{c\pi})=0$. $\mu_I^{c\pi}$ increases with magnetic field, which is qualitatively consistent with the conclusion of LQCD~\cite{lqcdb1} and model calculations~\cite{pib3}, and this means that magnetic field delays/disfavors the pion superfluid phase transition at finite isospin chemical potential. Physically, it can be understood in this way. Locating both the two constituent quarks at the lowest Landau level are forbidden for charged pions due to its zero spin. According to the quark energy $E_f=\sqrt{p^2_3+2 n |Q_f B|+m_q^2}$, different electric charges of $u$ and $d$ quarks indicate different effective quark mass $\sqrt{2 n |Q_f B|+m_q^2}$ with finite magnetic field and zero momentum $p_3=0$. This mass difference plays the role of effective Fermi surface mismatch when $u$ quark and $\bar d$ quark form cooper pairs. The larger the magnetic field (mass difference) is, the more difficult to form pion superfluid becomes, and this leads to the increasing $\mu_I^{c\pi}$ in magnetic field.

Critical isospin chemical potential $\mu_I^{cc}$ for chiral restoration phase transition, see green dotted line in Fig.\ref{muicb}, is determined by the dynamical quark mass. At finite magnetic field, chiral restoration is a first order phase transition, and the quark mass jumps from a large value to a small value. It is noticeable that $\mu_I^{cc}$ and $\mu_I^{c\pi}$ are different from each other, except for one point at $eB=4.75m^2_\pi$, with $\mu_I^{cc}>\mu_I^{c\pi}$ at $eB<4.75m^2_\pi$ and $\mu_I^{cc}<\mu_I^{c\pi}$ at $eB>4.75m^2_\pi$.

The critical isospin chemical potential $\mu_I^{c\pi}$ is separated into two parts, denoted by the connecting point of red and black solid lines at $eB=4.75m^2_\pi$ in Fig.\ref{muicb}. For $eB<4.75m^2_\pi$, we observe that the critical isospin chemical potential is equal to the $\pi^+$ mass in magnetic field, with $\mu_I^{c\pi}=M_{\pi^+}=m_{\pi^+}(B,T=\mu_B=\mu_I=0)$, as shown by the overlap between the black solid line and blue dashed line in Fig.\ref{muicb}. This conclusion can be analytically proved, similar as the case without magnetic field~\cite{model9}. At $T=0$, the Fermi-Dirac distribution $f(x)$ becomes a Heaviside step function $\theta(-x)$. With fixed magnetic field, we solve a constant quark mass $m_q(B,T=\mu_B=0,\mu_I)=m_q(B,T=\mu_B=\mu_I=0)$ from gap equation (\ref{gap}), before the chiral restoration happens. And by straightforward comparison of gap equation (\ref{gap}) and pole equation (\ref{pip}), a linearly decreasing $\pi^+$ mass is obtained $m_{\pi^+}(B,T=\mu_B=0,\mu_I)=M_{\pi^+}- \mu_I$. Applying the Goldstone's theorem, the critical isospin chemical potential $\mu_I^{c\pi}$ for pion superfluid is determined by the condition $m_{\pi^+}(B,T=\mu_B=0,\mu_I^{c\pi})=0$. Therefore, we solve $\mu_I^{c\pi}=M_{\pi^+}$. At $eB=4.75m^2_\pi$, both the pion superfluid phase transition and the chiral restoration phase transition happen at the same critical isospin chemical potential $\mu_I^{c\pi}=\mu_I^{cc}$. Since chiral restoration is a first order phase transition, associated with the quark mass jump. It leads to the discontinuous $\mu_I^{c\pi}$ for pion superfluid phase transition, as shown by the different slope of black and red lines around $eB=4.75m^2_\pi$. For $eB>4.75m^2_\pi$, no such analytical derivations are available and we should rely on the numerical calculations. The critical isospin chemical potential $\mu_I^{c\pi}$ is deviated from $M_{\pi^+}$, although they both increase in magnetic fields. With stronger magnetic field, the deviation becomes larger.

\begin{figure}[hbt]
\centering
\includegraphics[width=7.5cm]{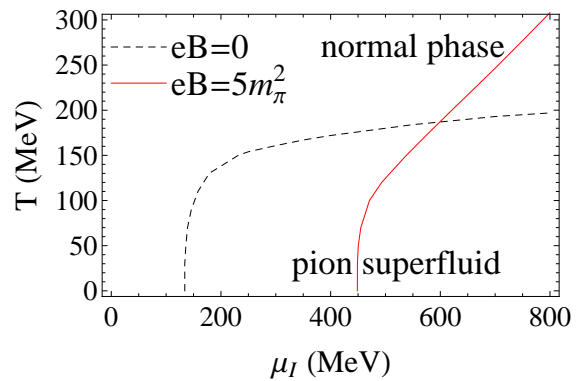}
\caption{Pion superfluid phase diagram in $\mu_I-T$ plane with $\mu_B=0$ and fixed magnetic field. The black dashed line is for $eB/m^2_\pi=0$, and red solid line for $eB/m^2_\pi=5$. } \label{tmui}
\end{figure}

We now turn on the temperature effect and depict the pion superfluid phase diagram in $\mu_I-T$ plane with $\mu_B=0$ and fixed magnetic field $eB/m^2_\pi=0$ (black dashed line) and $eB/m^2_\pi=5$ (red solid line) in Fig.\ref{tmui}. The phase transition line determined by the massless $\pi^+$ meson divides the $\mu_I-T$ plane into two regions. The pion superfluid phase is located in high isospin chemical and low temperature region, and the quarks are in normal phase for low isospin chemical potential and/or high temperature region. With increasing temperature, the quark thermal motion becomes strong. It prohibits the quark pairing and leads to the phase transition from pion superfluid phase to normal phase. The critical temperature increases with isospin chemical potential. Comparing with vanishing magnetic field case, the pion superfluid phase is retarded to higher isospin chemical potential, and it survives in higher temperature under finite magnetic field.

\begin{figure}[hbt]
\centering
\includegraphics[width=7.5cm]{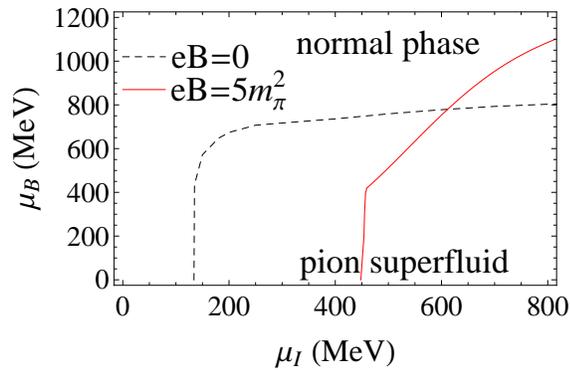}
\caption{Pion superfluid phase diagram in $\mu_I-\mu_B$ plane with $T=0$ and fixed magnetic field. The black dashed line is for $eB/m^2_\pi=0$, and red solid line for $eB/m^2_\pi=5$.} \label{mubmui}
\end{figure}

Fig.\ref{mubmui} is the phase diagram in $\mu_I-\mu_B$ plane with $T=0$ and fixed magnetic field. The black dashed line is for $eB/m^2_\pi=0$, and red solid line for $eB/m^2_\pi=5$. Pion superfluid phase locates in high isospin chemical potential and low baryon chemical potential region. In the low isospin chemical potential and/or high baryon chemical potential region, quarks are in normal phase. At zero baryon chemical potential, the $u$ quark and $\bar d$ quark form coherent pairs and condensate on a uniform Fermi surface, as $\mu_I>\mu_I^{c\pi}$. When the baryon chemical potential is switched on, there appears a Fermi surface mismatch between the $u$ quark and $\bar d$ quark, and it causes the phase transition from pion superfluid phase to normal phase. The critical baryon chemical potential increases with isospin chemical potential. With stronger magnetic field, the pion superfluid phase happens at higher isospin chemical potential and survives at higher baryon chemical potential. It should be mentioned that even in large baryon chemical potential case, we still neglect the color superconductor phase. The competition between color superconductor and pion superfluid in $\mu_I-\mu_B$ plane will be studied elsewhere.

Magnetic field effect on pion superfluid phase transition is studied in frame of a Pauli-Villars regularized NJL model. Instead of directly dealing with charged pion condensate, we apply the Goldstone's theorem (massless Goldstone boson $\pi^+$) to determine the onset of pion superfluid phase. Seriously taking into account the breaking of translational invariance, the charged pion propagator is constructed at finite magnetic field, temperature and chemical potential, and the $\pi^+$ mass and pion superfluid phase diagram are obtained. At weak magnetic field and vanishing temperature and baryon chemical potential, it is analytically proved that the critical isospin chemical potential $\mu^{c\pi}_I$ is equal to the $\pi^+$ mass in magnetic field, $\mu^{c\pi}_I=M_{\pi^+}$. Under external magnetic field, the pion superfluid phase is retarded to higher isospin chemical potential, and can survive at higher temperature and higher baryon chemical potential.\\

\noindent {\bf Acknowledgement:}
The work is supported by the NSFC Grant 11775165 and Fundamental Research Funds for the Central Universities.

\end{document}